\documentclass{aa}  
\usepackage{graphicx}
\usepackage{txfonts}
\begin{document}


\title{Neutrino transport and  hydrodynamic stability of rotating 
proto-neutron stars}
\author{V.~ Urpin$^{1,2)}$}
\offprints{Vadim.Urpin@uv.es}
\institute{$^{1)}$ Departament de Fisica Aplicada, Universitat d'Alacant,
           Ap. Correus 99, 03080 Alacant, Spain \\
           $^{2)}$ A.F.Ioffe Institute of Physics and Technology and Isaac
           Newton Institute of Chile, Branch in St.Petersburg,
           194021 St.Petersburg, Russia}

\date{today}

\abstract
{}
{We study stability of differentially rotating non-magnetic
proto-neutron stars.} 
{Linear stability properties are considered taking into account neutrino 
transport.} 
{If neutrino transport is efficient, the 
star can be subject to a diffusive instability that can occur even in the 
convectively stable region. The instability arises on the time-scale 
comparable to the time-scale of thermal diffusion.} 
{Hydrodynamic motions 
driven by the instability can lead to anisotropy in the neutrino flux 
since the instability is suppressed near the equator and rotation axis.}      

\maketitle

\section{Introduction}

Newly born neutron stars are subject to various hydrodynamic and 
hydromagnetic instabilities shortly after core collapse. These 
instabilities can play an important role in enhancing neutrino 
luminosities and increasing the energy deposition efficiency. Convection 
is probably among the best-studied instabilities in proto-neutron stars 
(PNSs) where it can be driven by both lepton and entropy gradients (see, 
e.g., Epstein 1979; Arnett 1987; Burrows \& Lattimer 1988). The 
development of negative entropy and lepton gradients is common to many 
simulations of core collapse supernovae (Bruenn \& Mezzacappa 1994; 
Bruenn, Mezzacappa \& Dineva 1995; Thompson et al. 2005) and evolutionary 
models of PNSs (Burrows \& Lattimer 1986; Keil \& Janka 1995; Sumiyoshi 
et al. 1995; Pons et al. 1999; Dessart et al. 2006). Numerical simulations
also indicate the presence of instabilities in PNSs, however, conclusions 
regarding the efficiency of turbulent transport and the duration of 
unstable phase depend on the code used and can differ significantly; 
compare, for instance, the simulations by Keil et al. (1996) and Mezzacappa
et al. (1998). 

The nature of instabilities arising in PNSs has been considered by a 
number of authors (Grossman et al. 1993; Bruenn \& Dineva 1996; Miralles 
et al. 2000, 2002). The true stability criteria for non-magnetic and 
non-rotating PNSs (Miralles et al. 2000) indicate the presence of two 
essentially different instabilities with the convectively 
unstable region surrounded by the neutron-finger unstable region, the 
latter typically involving a larger portion of the stellar material. Due 
to the cooling, the temperature and lepton gradients are progressively 
reduced, and both instabilities disappear completely after $\sim 30-40$ 
s. The growth time of convective and neutron-finger instabilities differs 
substantially, and therefore the efficiency of turbulent transport in the 
convective and neutron-finger unstable zones can be different as well. 
This difference is crucial for many MHD-phenomena in PNSs, 
such as turbulent dynamo action (see, e.g., Bonanno et al. 2003, 2005) 
and transport of the angular momentum. Note that fast rotation of PNSs 
can substantially modify convection and make convective motions 
anisotropic and constrained to the polar regions (Fryer \& Heger 2000, 
Miralles et al. 2004). This mechanism is a natural way of creating 
anisotropic energy and momentum transport by convective motions, which 
only requires that the angular velocity be of the order of the 
Brunt-V\"ais\"al\"a or leptonic buoyant frequences. 

The hydrodynamic stability properties of proto-neutron stars, however,
can be much more complex than is believed at the moment. Recently,
for example, Bruenn et al. (2004) have argued that a new doubly diffusive 
instability can occur in an extensive region below the neutrinosphere,
which the authors refer to as ``lepto-entropy fingers''. This instability
is driven mainly by the entropy equilibration due to a lepton fraction
difference and lepton equilibration that is in turn caused by the entropy 
difference and that may play an important role in enhancing the neutrino 
emission.

The possible presence of the magnetic field and differential rotation in a
core collapse supernovae favours a number of other instabilities that can 
also be important in enhancing anisotropic turbulent transport. For
example, the magnetorotational instability in the context of core collapse
was first considered by Akiyama et al. (2003). The authors argue that 
the instability must inevitably occur in core collapse and that it has the
capacity to produce fields that are strong enough to affect, if not 
cause, the explosion. Thompson et al.(2005) constructed one-dimensional 
models, including rotation and magnetic fields, to study the mechanism of 
energy deposition. They explore several mechanisms for viscosity and argue 
that the turbulent viscosity caused by the magnetorotational instability is 
the most effective. The nonaxisymmetric magnetorotational instability in 
PNSs has been considered by Masada et al. (2006a). The authors have 
obtained the criterion and growth rate of instability and argue that 
the nonlinear evolution may lead to enhancement of the neutrino 
luminosity. The effect of neutrino radiation on this instability was then 
considered by Masada et al. (2006b). Neutrino transport is rather fast in 
PNSs and can essentially modify the standard magnetorotational instability. 

In the present paper, we consider the effect of neutrino transport on the 
stability of differential rotation in PNSs, which can appear for a number 
of reasons. From theoretical modelling 
and simple analytic considerations, it is commonly accepted that the core 
collapse of a rotating progenitor leads to differential rotation for 
newly-born neutron stars (Zwerger \& M\"uller 1997; Rampp et al. 1998; 
Liu 2002; Dimmelmeier et al. 2002; M\"uller et al. 2004), mainly due to 
conservation of the angular momentum during collapse. It is also possible
that PNSs are differentially rotating due to turbulent transport of the
angular momentum during the convective phase. The boundaries of the
convective zone move inward when the PNS cools down and, in the region
where convection stops, the angular velocity profile will evolve under 
the influence of shear viscosity and, what is more likely, due to 
development of various instabilities caused by differential rotation. 

We show that this rotation can be unstable if it is accompanied by 
sufficiently rapid diffusion of heat and lepton number and if the angular 
velocity depends on the vertical coordinate. Contrary to the magnetorotational
instability, the instability considered in this paper does not require the 
presence of the magnetic field. Turbulent motions caused by the instability 
can lead to anisotropic turbulent transport in convectively stable regions. 
The instability can be important when considering various dynamo models, the 
transport of energy, and the angular momentum in PNSs.  

The paper is organised as follows. In Section 2, we consider the basic
equations governing the instability in differentially rotating PNSs and
derive the dispersion equation. The stability criterion and the growth
rate of instability are treated in Section 3. A discussion of our
results is presented in Section 4.

\section{Dispersion equation}

We now consider the PNS rotating with the angular velocity $\Omega =
\Omega(s,z)$, 
where $s$, $\varphi$, and $z$ are cylindrical coordinates. In the 
unperturbed state, the PNS is assumed to be in hydrostatic equilibrium,
\begin{equation}
\frac{\nabla p}{\rho} = \vec{G} \;, \;\;\;
\vec{G} = \vec{g} + \Omega^{2} \vec{s}
\end{equation}
where $\vec{g}$ is the gravity and $\vec{e}_{s}$ is the radial unit 
vector. We assume that the matter inside a PNS is in chemical equilibrium; 
thus, the pressure $p$ is generally a function of the density $\rho$, 
temperature $T$, and lepton fraction $Y=(n_{e} + n_{\nu})/n$, with 
$n_{e}$ and $n_{\nu}$ the number densities of electrons and 
neutrinos, respectively, and $n=n_{p}+n_{n}$ is the number density of 
baryons. Therefore, we have $\nabla \ln p =(\partial \ln p/\partial 
\ln \rho)_{TY} (\nabla \ln \rho + \beta \nabla \ln T + \delta \nabla Y)$, 
where $\beta$ and $\delta$ are the coefficients of thermal and chemical 
expansion; $\beta = - (\partial \ln \rho/\partial \ln T)_{pY}$, $\delta 
= - (\partial \ln \rho/ \partial Y)_{pT}$. Then, taking the curl of 
Eq.~(1), we obtain
$$
\vec{G} \times \left( \beta \frac{\nabla T}{T} + \delta \nabla Y \right)
= - \vec{e}_{\varphi} s \frac{\partial \Omega^2}{\partial z}.
$$
If $\Omega$ depends on $z$, hydrostatic equilibrium can be reached only
if gradients of the temperature or lepton fraction have components
perpendicular to $\vec{G}$. If $g \gg s \Omega^2$, then the variations
of $T$ and $Y$ on a spherical surface $r=$const can be estimated as  
$$
\max \left( \frac{\delta T_{\perp}}{T}, \delta Y_{\perp} \right) \sim
\frac{1}{\beta + \delta} \frac{s^2}{g} \frac{\partial 
\Omega^2}{\partial z}.
$$
Hydrostatic equilibrium can be satisfied if variations are small, or if 
$s \Omega^2/g (\beta + \delta) \ll 1$ (we assume that $\partial \Omega/
\partial z \sim \Omega/s$). Typically, $\beta + \delta \sim 0.1$ or 
larger in PNSs (see Miralles et al. 2002) and, hence, hydrostatic 
equilibrium can be reached for PNSs with $\partial \Omega/ \partial z 
\neq 0$ if the rotation period is longer than $\sim 1-2$ ms. Most likely, 
the majority of PNSs satisfies this requirement and, therefore, the 
unperturbed state with $\Omega(s, z)$ is justified for these PNSs.

We consider axisymmetric short wavelength perturbations with the 
space-time dependence $\exp(\gamma t - i \vec{k} \cdot \vec{r})$ where 
$\vec{k}= (k_{s}, 0, k_{z})$ is the wavevector, $|\vec{k} \cdot \vec{r}| 
\gg 1$. Small perturbations will be indicated by subscript 1, whilst 
unperturbed quantities will have no subscipt. Since the growth time of the 
instability in PNSs is typically longer than the period of a sound wave 
with the same wavelength, we use the Boussinesq approximation. Then, 
the linearized momentum and continuity equations governing the behaviour 
of small perturbations read as  
\begin{eqnarray}
\gamma \vec{v}_{1} + 2 \vec{\Omega} \times \vec{v}_{1} +
\vec{e}_{\varphi} s (\vec{v}_{1} \cdot \nabla \Omega) =
\frac{i \vec{k} p_{1}}{\rho} + \vec{G} \frac{\rho_{1}}{\rho} 
\nonumber \\
- \nu k^{2} \vec{v}_{1},
\end{eqnarray}
\begin{equation}
\vec{k} \cdot \vec{v}_{1} = 0 \;, 
\end{equation}
where $\vec{v}_{1}$, $p_{1}$, and $\rho_{1}$ are perturbations of the 
velocity, pressure, and density, respectively, and $\nu$ is the kinematic 
viscosity and $\vec{e}_{\varphi}$ the unit vector in the azimuthal 
direction. Since in the Boussinesq approximation, the perturbations of 
pressure are negligible, the perturbations of density can be expressed 
in terms of the perturbations of temperature and lepton fraction,
\begin{equation}
\rho_{1} \approx - \rho \left( \beta \frac{T_{1}}{T} + \delta Y_{1} 
\right).
\end{equation}

During the initial evolution, the PNS is opaque to neutrinos, and the 
neutrino transport can be treated in the diffusion approximation (Imshenik
\& Nadezhin 1972). Then, the linearized thermal balance and diffusion 
equations read as
\begin{equation}
\frac{\dot{T}_{1}}{T} - {\bf v}_{1} \cdot  \frac{\Delta\nabla T}{T} =
\kappa_{T} \frac{\Delta T_{1}}{T} + \kappa_{Y} \Delta Y_{1}, 
\end{equation}
\begin{equation}
\dot{Y}_{1} + {\bf v}_{1} \cdot \nabla Y =  
\lambda_{T} \frac{\Delta T_{1}}{T} + \lambda_{Y} \Delta Y_{1}\;
\end{equation}
where $\lambda_{T}$, $\lambda_{Y}$, $\kappa_{T}$, and $\kappa_{Y}$ are 
the corresponding kinetic coefficients (see Miralles et al. 2000). 
We denote the superadiabatic temperature gradient as
\begin{equation}
\Delta\nabla T= 
\left( \frac{\partial T}{\partial p} \right)_{s,Y} \nabla p
-  \nabla T. 
\end{equation}. 

The dispersion equation corresponding to Eqs.~(2)-(6) is
\begin{equation}
\gamma^{4} + a_{3} \gamma^{3} + a_{2} \gamma^{2} + a_{1} \gamma 
+ a_{0} = 0,
\end{equation}
where 
\begin{eqnarray}
a_{3} = 2 \omega_{\nu} + \omega_{T} + \omega_{Y},  \nonumber  \\
a_{2} = \omega_{T} \omega_{Y} - \omega_{TY} \omega_{YT} +
2 \omega_{\nu} (\omega_{T} + \omega_{Y}) + \omega_{\nu}^{2} + q^{2}
\nonumber \\
- (\omega_{g}^{2} + \omega_{L}^{2}),  \nonumber  \\
a_{1} = 2 \omega_{\nu} (\omega_{T} \omega_{Y} - \omega_{TY} \omega_{YT})
+ (\omega_{T} + \omega_{Y}) (\omega_{\nu}^{2} + q^{2}) 
\nonumber \\
- \omega_{\nu} (\omega_{g}^{2} + \omega_{L}^{2}) 
- \omega_{g}^{2} \left( \omega_{Y} - \frac{\delta}{\beta} \omega_{YT} 
\right) - \omega_{L}^{2} \left( \omega_{T} - \frac{\beta}{\delta} 
\omega_{TY} \right).
\nonumber \\
a_{0} = - \omega_{\nu} \left[ \omega_{g}^{2} \left( \omega_{Y} - 
\frac{\delta}{\beta} \omega_{YT} \right) + \omega_{L}^{2} \left( 
\omega_{T} 
- \frac{\beta}{\delta} \omega_{TY} \right) \right]
\nonumber \\
+ (\omega_{\nu}^{2} + q^{2}) (\omega_{T} \omega_{Y} - 
\omega_{TY} \omega_{YT}).
\nonumber 
\end{eqnarray}
In these expressions, we introduced the characteristic frequencies
\begin{eqnarray}
q^{2} = \frac{k_{z}}{s^{3} k^{2}} \left( k_{z}
\frac{\partial}{\partial s}  -  k_{s} \frac{\partial}{\partial z} \right) 
(s^{4} \Omega^{2}), &\;& \omega_{\nu} = \nu k^{2}, 
\nonumber \\ 
\omega_{T} = \kappa_{T} k^{2}, &\;&  \omega_{Y} = \lambda_{Y} k^{2},   
\nonumber \\
\omega_{YT} = \lambda_{T} k^{2}, &\;& \omega_{TY} = \kappa_{Y} k^{2}, 
\nonumber \\
\omega_{g}^{2} = - \beta \vec{A} \cdot \frac{\Delta\nabla T}{T}, &\;&
\omega_{L}^{2} = \delta \vec{A} \cdot \nabla Y ,        \nonumber
\end{eqnarray}
where $\vec{A} = \vec{G} - \vec{k} ( \vec{k} \cdot \vec{G})/k^{2}$. 
The quantities $\omega_{\nu}$, $\omega_{T}$, and $\omega_{Y}$ are the
inverse time scales of the dissipation of perturbations due to viscosity, 
thermal conductivity, and diffusivity, respectively; $\omega_{YT}$
characterises the rate of thermodiffusion, and $\omega_{TY}$ describes 
the rate of heat transport due to the chemical inhomogeneity;
$\omega_{g}$ is the frequency of the buoyant wave; $\omega_{L}$ 
characterises the dynamical time scale of the processes associated 
with the lepton gradient, and $q^{2}$ describes the influence of differential 
rotation on stability. 

If rotation is negligible ($q^{2} \approx 0$), then Eq.~(8) reduces
to a cubic one,
\begin{equation}
\gamma^{3} + b_{2} \gamma^{2} + b_{1} \gamma + b_{0} = 0,
\end{equation}
where
\begin{eqnarray}
b_{2} = \omega_{\nu} + \omega_{T} + \omega_{Y},  \nonumber  \\
b_{1} = \omega_{T} \omega_{Y} - \omega_{TY} \omega_{YT} +
\omega_{\nu} (\omega_{T} + \omega_{Y}) - (\omega_{g}^{2} +
\omega_{L}^{2}),  \nonumber  \\
b_{0} = 
- \omega_{g}^{2} \left( \omega_{Y} - \frac{\delta}{\beta} \omega_{YT} 
\right) 
- \omega_{L}^{2} \left( \omega_{T} - \frac{\beta}{\delta} \omega_{TY} 
\right)
\nonumber \\
+ \omega_{\nu} (\omega_{T} \omega_{Y} - \omega_{TY} \omega_{YT}).
\nonumber  
\end{eqnarray}
This equation is equivalent to the dispersion equation (17) derived by 
Miralles et al.(2000).

\section{Criteria and the growth rate of instability}

Equation (8) describes four essentially different modes that exist in a 
differentially rotating PNS. The condition that one of the roots has a 
positive real part (unstable) is equivalent to requiring that at least 
one of the following inequalities be satisfied:
\begin{eqnarray}
a_{3} < 0, \\
a_{0} < 0, \\
a_{3} a_{2} - a_{1} < 0, \\
a_{1} (a_{3} a_{2} - a_{1}) - a_{3}^{2} a_{0} < 0  
\end{eqnarray} 
(see, e.g., Aleksandrov et al. 1963). The quantity $a_{3}$ is 
positive, therefore condition (10) will never apply.

We have from condition (11)
\begin{eqnarray}
-  \nu \omega_{g}^{2} \left( \lambda_{Y} - \frac{\delta}{\beta}
\lambda_{T} \right) - \nu \omega_{L}^{2} \left( \kappa_{T} - 
\frac{\beta}{\delta} \kappa_{Y} \right)  
\nonumber \\
+ (q^{2} + \omega_{\nu}^2)( \kappa_{T} \lambda_{Y} - \kappa_{Y} 
\lambda_{T}) < 0. 
\end{eqnarray}
Note that, typically, the Prandtl number (the ratio of kinematic 
viscosity and thermal diffusivity) is small in PNSs (see, e.g., Thompson 
\& Duncan 1993; Masada et al. 2006b), and the stabilizing influence of 
stratification is reduced.

Substituting expressions for the coefficients $a_{0}$, $a_{1}$, $a_{2}$,
and $a_{3}$ into conditions (12)-(13), we obtain rather cumbersome 
criteria of instability. However, these criteria can be much simplified
if we take into account that the frequencies have a different order of
magnitude in PNSs. The dissipative frequencies $\omega_{T}$, $\omega_{Y}$,
$\omega_{TY}$, and $\omega_{YT}$ are approximately comparable to each 
other, and $\sim 1-10$ s$^{-1}$ for perturbations with the wavelength of 
the order of the density lengthscale. The dynamical frequencies, 
$\omega_{g}$ and $\omega_{L}$, are typically much higher at $\sim 10^3$ 
s$^{-1}$. Therefore, if analysing the conditions (12)-(13), we can restrict 
ourselves by terms of the highest order in dynamical frequencies and 
$q^{2}$ since the angular velocity of PNSs is very likely much higher than 
$1-10$ s$^{-1}$. The criteria of instability (12)-(13) read in the 
approximation as
\begin{eqnarray}
- \omega_{g}^{2} \left( \nu + \kappa_{T} + \frac{\delta}{\beta}
\lambda_{T} \right) - \omega_{L}^{2} \left( \nu + \lambda_{Y} + 
\frac{\beta}{\delta} \kappa_{Y} \right)  
\nonumber \\
+ 2 \nu q^{2} < 0, \\
- \omega_{g}^{2} \left( \nu + \lambda_{Y} - \frac{\delta}{\beta}
\lambda_{T} \right) - \omega_{L}^{2} \left( \nu + \kappa_{T} - 
\frac{\beta}{\delta} \kappa_{Y} \right)  
\nonumber \\
+ q^{2} (\kappa_{T} + \lambda_{Y}) < 0. 
\end{eqnarray}   
Conditions (15) and (16) represent neutron-finger and convective 
instability modified by rotation. If $\Omega =0$, these conditions 
agree with the criteria derived by Miralles et al. (2000). The effect 
of rotation is important for convective and neutron-finger instabilities 
if the angular velocity is comparable to the  Brunt-V\"ais\"al\"a or 
leptonic buoyant frequences. In this paper, we concentrate on 
condition (14), which represents a new instability in PNSs associated to 
differential rotation and neutrino transport.

We next consider the case of a moderately rapid rotation when  the PNS 
is far from the rotational distortion and $g \gg \Omega^{2}s$. 
Then, $\Delta \nabla T$, $\nabla Y$, and $\vec{g}$ are 
approximately radial, and 
\begin{equation}
\omega_{g}^{2} = - \beta \; \frac{k_{\theta}^{2}}{k^{2}} \; 
(\vec{G} \cdot \Delta \nabla T) = - \frac{k_{\theta}^{2}}{k^{2}} 
\omega_{BV}^{2}, 
\end{equation}
\begin{equation}
\omega_{L}^{2} = \delta \; \frac{k_{\theta}^{2}}{k^{2}}
(\vec{G} \cdot \nabla Y) = - \frac{k_{\theta}^{2}}{k^{2}}
\omega_{LO}^{2},
\end{equation}
where $\omega_{BV}^{2} = \beta ( \vec{g} \cdot \Delta \nabla T)$ is the 
Brunt-V\"ais\"al\"a frequency, and $\omega_{LO}^{2} = \delta (\vec{G} 
\cdot \nabla Y)$ is the characteristic frequency of leptonic buoyant 
oscillations; $k_{\theta}$ is the $\theta$-component of a wavevector, 
$k_{\theta}^{2}= k_{s}^{2} \cos^{2} \theta - 2 k_{s} k_{z} \cos
\theta \sin \theta + k_{z}^{2} \sin^{2} \theta$, $\theta$ is the
polar angle. Substituting expressionns (17)-(18) into condition (14),
we have
\begin{equation}
q^2 + \omega_{\nu}^2 + \frac{k_{\theta}^2}{k^2} \; \xi \; \omega_b^2 < 0,
\end{equation}
where
$$ 
\omega_b^2 =
\omega_{BV}^{2} + \omega_{LO}^{2}
\frac{\kappa_{T} - \beta \kappa_{Y} / \delta }{\lambda_{Y} - \delta 
\lambda_{T} /\beta}, \;\; 
\xi = \frac{\nu (\lambda_{Y} 
- \delta
\lambda_{T}/\beta)}{ \kappa_{T} \lambda_{Y} - \kappa_{Y} \lambda_{T}}.
$$
Here, $\omega_{b}$ is the characteristic buoyant frequency that is 
determined by the Bruni-V\"ais\"al\"a frequency and the frequency of 
leptonic buoyant oscillations.

Since $\kappa_{T} \lambda_{Y} - \kappa_{Y} \lambda_{T} >0$ from 
thermodynamics (see Pons et al. 1999), rotation-induced instability can 
occur in a stable stratification only if
\begin{equation}
q^2 = \frac{k_{z}}{s^{3} k^{2}} \left( k_{z}
\frac{\partial}{\partial s}  -  k_{s} \frac{\partial}{\partial z} \right) 
(s^{4} \Omega^{2}) < 0.
\end{equation}
This condition is satisfactory if either the specific angular momentum, $s^2 
\Omega$, decreases as the cylindrical radius increases or $\Omega$ depends
on $z$ such that $\partial \Omega /\partial z \neq 0$. 

Condition (19) depends on the direction of the wavevector and can be 
written as
\begin{equation}
F \equiv \omega_{\nu}^{2} + 
A \frac{k_{z}^{2}}{k^{2}} - B \frac{k_{s} k_{z}}{k^{2}} + 
C \frac{k_{s}^{2}}{k^{2}} < 0,
\end{equation}
where
\begin{eqnarray}
&& A = 4 \Omega^2 + s \Omega_{s}^{2} + \xi \omega_{b}^2 \sin^{2} \theta, 
\nonumber \\
&& B = s \Omega_{z}^{2} + \xi \omega_{b}^2 \sin 2\theta, \nonumber \\
&& C = \xi \omega_{b}^2 \cos^{2} \theta. \nonumber 
\end{eqnarray}
In these expressions, we denote
\begin{eqnarray}
\Omega_{s}^{2} = \frac{\partial \Omega^{2}}{\partial s} \;, \;\;
\Omega_{z}^{2} = \frac{\partial \Omega^{2}}{\partial z} \;.
 \end{eqnarray}

Since the dependence of $F$ on the direction of $\vec{k}$ is quadratic, 
we can obtain that the relative maximum of $F$ corresponds to
\begin{equation}
\frac{k_{z}^{2}}{k^{2}} = \frac{1}{2} 
\left[ 1 \pm \sqrt{\frac{(A-C)^{2}}{(A-C)^2 + B^2}} \right]~.
\label{phimax}
\end{equation}
The maximum value of $F$ corresponding to these $k_{z}^{2}/k^{2}$
yields the following condition of instability: 
\begin{equation}
2 \omega_{\nu}^{2} + A + C \pm
\sqrt{ B^{2} + (A-C)^{2}} < 0. 
\end{equation}
The two conditions for instability then follow from the above expression:
\begin{equation}
A+C+2 \omega_{\nu}^2 = \kappa^2 + \xi \omega_{b}^2
+ 2 \omega _{\nu}^2 < 0, 
\end{equation}
or
\begin{equation}
B^2 + (A - C)^2 > (A+C + 2 \omega_{\nu}^2 )^2,
\end{equation}
where $\kappa^2 = 4 \Omega^2 + s \Omega_{s}^2$ is the epicyclic frequency.

Inequality (25) can be fulfilled only if differential rotation satisfies 
the Rayleigh criterion, $\partial (s^2 \Omega) / \partial s < 0$. This 
criterion requires differential rotation that decreases very rapidly 
with the cylindric radius and usually is not fulfilled in stars. 
Condition (26) can be rewritten as 
\begin{equation}
B^2 - 4 (A + \omega_{\nu})(C + \omega_{\nu}) > 0,
\end{equation}
or
\begin{eqnarray}
s^2 (\Omega_{z}^2)^2 + 2s \Omega_{z}^2 \xi \omega_{b}^2 \sin 2 \theta
- 4 \kappa^2 \xi \omega_{b}^2 -   \nonumber \\
4 \omega_{\nu}^2 (\kappa^2 + \xi \omega_{b}^2 +\omega_{\nu}^2) > 0. 
\end{eqnarray}
If stratification is stable ($\xi \omega_{b}^2 > 0$) and rotation 
satisfies the Rayleigh stability criterion ($\kappa^2 > 0$), then the 
rotation-driven instability can occur only if $\partial \Omega / \partial
z \neq 0$. Generally, such rotation can often be achieved in PNSs. Note, 
however, that even if the angular velocity depends on $z$, the instability
does not arise near the rotation axis where $s$ is small and near the 
equator where $\partial \Omega / \partial z = 0$.

Since differential rotation is likely to be very strong in PNSs, we can 
estimate
$s \Omega_{z}^2 \sim \Omega^2$. The ``viscous'' frequency is negligible in
Eq.(28) if the wavelength of perturbations $\lambda = 2 \pi/ k$ satisfies 
the condition 
\begin{equation}
\lambda > \lambda_{cr} = 2 \pi \sqrt{\nu / \Omega}.
\end{equation}
Since $\nu \sim 10^8 - 10^9$ cm$^2$/s in PNSs (see, e.g., Thompson \& 
Duncan 1993; Masada et al. 2006), we can estimate $\lambda_{cr}$ as 
$10^4$ cm, which allows a comfortable range of unstable wavelengths. For 
$\lambda > \lambda_{cr}$, the necessary condition of instability is 
\begin{equation}
(s \Omega_{z}^2)^2 + 2 s \Omega_{z}^2  \xi \omega_{b}^2 
\sin 2 \theta   - 4 \kappa^2 \xi \omega_{b}^2 > 0. 
\end{equation}
If stratification is stable ($\xi \omega_b^2 > 0$), this inequality is 
equivalent to the conditions
\begin{eqnarray}
s \Omega_z^2 >  \xi \omega_b^2 \sin 2 \theta + \sqrt{ \xi^2 \omega_b^4 
\sin^2 2 \theta + 4 \kappa^2 \xi \omega_b^2} , \\
s \Omega_z^2 <  \xi \omega_b^2 \sin 2 \theta - \sqrt{ \xi^2 \omega_b^4 
\sin^2 2 \theta + 4 \kappa^2 \xi \omega_b^2} .
\end{eqnarray}  
The first condition requires a large positive radial gradient of 
$\Omega^2$ that seems to be unlikely in PNS. On the contrary, the second 
condition can be satisfied if the vertical gradient of the angular 
velocity is negative. If rotation is sufficiently fast and $\kappa^2 \sim 
\Omega^2 \gg \xi \omega_b^2$, condition (32) yields
\begin{equation}
\frac{|s \Omega_{z}^2|}{\sqrt{\kappa^2}} \sim \Omega \gg
2 \sqrt{\xi} \omega_{b}.
\end{equation}
Since  $\xi \sim \nu / \kappa_T \ll 1$, condition (33) can be fulfilled
in many PNSs even if $\Omega < \omega_b$.

Many studies model rotation of the collapsing core by the angular velocity 
profile that depends on the spherical radius $r$ alone, $\Omega = \Omega(r)$ 
(see, e.g., Akiyama et al. 2003; Kotake et al. 2004; Thompson et al. 2005; 
Sawai et al. 2005). Such a shellular rotation can be 
justified if the progenitor rotates with the angular velocity that 
depends on $r$ alone (M\"onchmeyer \& M\"uller 1989). In this case, 
conservation of the angular momentum in the course of collapse leads 
straightforwardly to a shellular rotation of the collapsing core at the 
beginning of evolution, at least (see, e.g., Akiyama et al. 2003).  
For shellular rotation, we have $\Omega_{s}^2 = \Omega_{r}^2 \sin 
\theta$ and $\Omega_{z} = \Omega_{r}^2 \cos \theta$ where $\Omega_r^2 
= d \Omega^2/dr$. Then, the criterion of instability reads
\begin{eqnarray}
\frac{|r \Omega_{r}^2|}{\sqrt{\kappa^2}} \sin 2 \theta \sim \Omega \gg
4 \sqrt{\xi} \omega_{b}.
\end{eqnarray}
In this case, the instability occurs neither near the rotation axis nor 
near the equator but is strongly constrained to the polar angle $\theta 
\sim \pi /4$.  

The growth rate of instability can be estimated from Eq.~(8) if we
assume that $\omega_b \sim \omega_{BV} > \Omega$ and condition
(30) is satisfied. Then, by making use of the perturbation procedure, we 
obtain for the unstable root, with accuracy in the lowest order in $\nu$ 
and $\Omega^2$,
\begin{equation}
\gamma \approx - \frac{a_0}{a_1} \approx - \frac{k^4 (\kappa_T \lambda_Y
- \kappa_Y \lambda_T)}{k_{\theta}^2 \omega_b^2 (\lambda_Y - 
\delta \lambda_Y/
\beta)} \left( \frac{k_{\theta}^2}{k^2} \xi \omega_b^2 +q^2 + 
\omega_{\nu}^2
\right).
\end{equation}
This root is positive (unstable) if criterion (19) is satisfied. Under 
conditions (29) and (33), expression (35) yields the following estimate
for the growth rate of perturbations with $k_{\theta} \sim k$:
\begin{equation} 
\gamma \sim \omega_T \; \frac{\Omega^2}{\omega_b^2}.
\end{equation}
Perturbations with small $k_{\theta}$ (that corresponds to $k \approx 
k_r$) can grow even faster. In this case, we also should keep in $a_1$  
the terms proportional to $q^2$ and obtain that $\gamma \sim \omega_T$. 
The growth rate of instability is rather high, particularly, for 
perturbations with a relatively short wavelength (but still satisfying 
condition (29)). Generally, $\gamma$ can be comparable to the growth 
rate of other diffusive instabilities in PNSs, such as 
the neutron-finger instability. 

\section{Discussion}

We have derived the criteria and growth rate of the rotation-driven 
instability in non-magnetic PNSs. This instability is associated with 
differential rotation and neutrino transport and can even arise in 
convectively stable regions. This instability is representative
of a wide class of the so-called doubly diffusive instabilities known in
stellar hydrodynamics (see, e.g., Acheson 1978). In fact, this instability
is the analogue of the well-known Goldreich-Schubert-Fricke instability 
for the case of neutrino transport (see Goldreich \& Schubert 1968; 
Fricke 1969). The criterion of the Goldreich-Schubert-Fricke instability 
in ordinary stars is $\partial \Omega / \partial z \neq 0$. However,
neutrino transport is qualitatively different from the radiative one 
even in diffusive approximation because, apart from transporting heat, 
it also involves transport of the lepton number (Eqs.~(5) and (6)), which
is important for the pressure balance. That is the reason the 
criterion of instability in PNSs (30) is so different from the 
standard condition $\partial \Omega / \partial z \neq 0$ derived by 
Goldreich \& Schubert (1968) and Fricke (1969). In PNSs, however, the 
instability also occurs only if the angular velocity depends on the 
vertical coordinate. 

During the early evolution stage, a rotation profile can be rather 
complex in PNSs and, therefore, the necessary condition of instability (33) 
can be fulfilled. For example, 2D calculations of the early rotation 
evolution of PNSs by Ott et al. (2006) indicate the presence of differential 
rotation in PNSs at 200 ms after bounce. Differential rotation is moderately
strong in the inner region of a PNS with $r<10$ km, where $\Omega$ varies
by a factor 1.5-3 depending on the model. In the outer region ($r>10$ km),
differential rotation is substantially stronger. 

The growth rate of the considered instability is small compared to that 
of convection; nevertheless, it is rather high and, in general, can be 
comparable to the growth rate of other diffusive instabilities such as 
the neutron-finger one. For example, even if rotation is slow and 
$\Omega \sim 0.1 \omega_b$, the growth time of perturbations with a 
lengthscale close to the pressure scale-height is $\sim 1$ s. 
Perturbations with a shorter wavelength grow much faster. Most likely, 
then, the unstable motions reach saturation in PNSs. The turbulent 
velocity in saturation can be estimated by making use of the standard 
mixing-length approximation (see, e.g., Schwarzschild 1958). In the case 
of diffusive instabilities, this approximation agrees very well with 
numerical simulations (see Arlt \& Urpin 2004). Then, the 
turbulent velocity with the lengthscale $\lambda$ is given by 
\begin{equation}
v(\lambda) \sim \lambda \gamma(\lambda) \sim \frac{\kappa_T}{\lambda} \;
\frac{\Omega^2}{\omega_b^2},
\end{equation}    
where $\gamma(\lambda)$ is the growth rate of perturbations with the 
wavelength $\lambda$ obtained from the linear theory of instability. If 
$\Omega \sim 0.1 -1 \; \omega_b$, then the velocity in the largest 
scale comparable to the pressure scale-height is $\sim 10^5 - 10^7$ cm/s. 
Such motions can essentially enhance, for example, the angular momentum 
transport that is important for the dynamo action in PNSs. The 
corresponding coefficient of turbulent diffusion is
\begin{equation}
\nu_T \sim \kappa_T \; \frac{\Omega^2}{\omega_b^2},
\end{equation}
and it can be comparable to the coefficient of heat diffusion that is 
largest among the diffisive coefficients in PNSs. Note that turbulent 
transport caused by the instability should be substantially anisotropic
since it cannot arise in regions close to the equator and 
rotation axis. Such anisotropic transport can perhaps lead to an 
anisotropic enhancement in the neutrino flux during the initial second 
following the core-bounce.

We have considered the instability of only short wavelength perturbations
but, most likely, perturbations with a wavelength of the order of the
pressure scale-heigh are unstable as well. Therefore, the 
instability can be the reason for a large-scale overturn of the core region 
with a high energy density. This overturn can generally enhance 
the neutrino flux and make it anisotropic, which is crucial for the 
explosion mechanism since even a slight modification in the neutrino 
cooling/heating efficiency changes it qualitatively (Dessart et al. 
2006). This is particularly important because the instability operates 
in the regions where convection is suppressed and cannot influence the 
neutrino transport. For example, according to Dessart et al. (2006), 
convection at $t < 1$ s occurs in the region between 10 and 30 km, whereas 
the innermost region with the highest energy content is stable. The 
rotation-induced instability in the core could lead to motions that 
transport heat and lepton number to the outer region. The important 
feature of the instability is that it produces large-scale 
departures from the spherical symmetry. This instability cannot arise 
near the equator and rotation axis and, therefore, turbulent transport 
is suppressed in those regions.     

Hydrodynamic motions induced by the instability can also contribute to 
the gravitational wave signals from the core-collapse supernovae. As 
recognised by M\"uller et al. (2004), the dominant contribution 
to the gravitational wave signal is not produced by the core bounce 
itself but instead by the neutrino-driven convection in the postshock 
region and the Ledoux convection inside the deleptonizing proto-neutron 
star; note, however, that Fryer et al. (2004) argue the opposite based on
their results. Because of these contradictory results, the influence of 
convective motions on the gravitational wave signal seems to be an open 
issue. It is possible, however, that motions caused by the rotational 
instability, despite being slower than convective motions, can produce 
a significant gravitational signal, because the instability arises in 
the dense inner region of the 
PNS with the radius $<10$ km where convection does not occur during the 
first second after the bounce.

\section*{Acknowledgement}

This work has been supported by the research grant from Generalitat 
Valenciana. The author thanks the University of Alicante for hospitality.

\end{document}